\documentstyle[graphicx,multicol,prl,aps,epsf]{revtex}
\begin{document}
\tightenlines
\title{Resonant electron transmission through a 
finite quantum spin chain}
\author{Y. Avishai$^{1,2}$ and Y. Tokura$^{1}$}
\address{$^{1}$ NTT Basic Research Laboratory, 3-1 Morinosato, 
Wakamiya, 243-01 Japan\\
$^2$ Ilze Katz Center and Department of Physics, Ben-Gurion 
University of the Negev,
Beersheva, Israel
}
\maketitle

\begin{abstract}
Electron transport in a finite one dimensional 
quantum spin chain (with ferromagnetic exchange) is studied within 
an $s-d$ exchange Hamiltonian. 
Spin transfer 
coefficients strongly depend on the sign of 
the $s-d$ exchange constant. For a ferromagnetic  coupling, 
they exhibit a novel resonant pattern, reflecting
the salient features of the combined electron-spin system. 
Spin-flip processes are inelastic and feasible at finite voltage 
or at finite temperature. 
\end{abstract}
\begin{multicols}{2}
\narrowtext
\noindent
{\bf Motivation and scope}:
There is a growing interest in electronic devices 
which transport electron spin\cite{Dieter}
(together with its charge). 
Beside the Kondo physics \cite{Wiel}, electron transmission through 
a region containing a lattice of magnetic atoms 
(mainly a domain wall) is a focus
of experimental\cite{Bruno} \cite{Garcia} \cite{Oshima} \cite{Ono} 
and theoretical\cite{Imamura} \cite{Yamanaka} \cite{Tokura} investigations. 
In this context, the basic ingredient is an exchange interaction 
$g {\bf s} \cdot {\bf S}_{n}$ between the electron spin 
operator ${\bf s}$ and an atomic spin ${\bf S}_{n}$ localized at 
a point $x_{n}$, with an ``s-d'' 
coupling $g$. 
Most theoretical works treat the atomic spins  
either purely classically (as localized magnetic fields), 
or semi classically (the operator ${\bf S}_{n}$ is replaced by
its expectation value $<{\bf S}_{n}>$ in the ground state 
of the quantum spin system).  It is expected, however, that
the quantum nature of the magnetic system will be crucial at 
low temperatures. \\
In the present work, transmission of electrons 
through a finite one dimensional quantum 
spin $1/2$ chain (with ferromagnetic exchange)
is studied,
starting from an $s-d$ exchange
Hamiltonian. The many body 
Kondo problem is avoided by an application of a
magnetic field which removes the degeneracy of the 
ground state. It also 
enables cutting the spectrum of the 
spin system off at the 
one magnon level. 
The ensuing scattering formalism
can then be handled within a transfer matrix algorithm, 
in terms of which the spin transfer coefficients (STC) 
are calculated as function of the electron Fermi energy.
It is found that for ferromagnetic $s-d$ exchange, 
the STC display 
a novel resonant structure which manifests the richness 
of the combined electron spin chain system.\\
{\bf Hamiltonian}:
Consider electrons (mass $m$ and charge $-e$) 
moving in a one dimensional wire 
(along $x$) interacting with a
chain of quantum spins ({\em e.g.} magnetic atoms)  
${\bf S}_{n}$, ($n=1,2,\ldots N$,
${\bf S}_{N+1}={\bf S}_{1}$)
localized at points 
$x_{n}=na$ ($n=1,2,...N$),
subject to a perpendicular 
magnetic field ${\bf H}=H_{z}\hat{{\bf z}}$. 
Henceforth $a$ and 
$\hbar^{2}/2 m a^{2}$ are exploited such that 
all length and energy quantities are 
dimensionless. The Hamiltonian of the 
system $H=H_{e}+H_{s-d}+H_{S}$ contains 
an electronic part $H_{e}$, 
an electron-spin $s-d$ exchange interaction $H_{s-d}$ 
and a term $H_{S}$ controlling  
the isolated spin system.  The latter is,
\begin{eqnarray}
&& H_{S}=
-\sum_{n}(J {\bf S}_{n} \cdot {\bf S}_{n+1}+E_{Z}S_{nz}),
\label{Eq_HS}
\end{eqnarray}
where $J>0$ and $E_{Z}$ is the atomic
Zeeman energy. 
The spin Hamiltonian $H_{S}$ is projected 
on the subspace spanned by its $N+1$ lowest energy states 
$|k>$ ($k=0,1,\ldots N$).  
If the magnetic field 
is strong enough such that $E_{Z} > JS$ then, beside 
the (non degenerate) ground state 
$|k=0>$ (with all spins up along $\hat {{\bf z}}$  
and energy $E_{0}=-N(S^{2}J+E_{Z}S)$ and $\sum_{n}S_{nz}=\frac {N} 
{2}$), these are
the one magnon spin waves (OMSW) with $\sum_{n}S_{nz}=\frac {N} {2} -1$.
In terms of  states $|n> \equiv S_{n}^{-}|0>$, 
the OMSW and 
their corresponding energies are,
\begin{eqnarray}
&& |k>=\frac{1}{\sqrt N} 
\sum_{n=1}^{N}e^{i \frac {2 \pi} {N} kn}|n>, \nonumber \ \ \ 
\mbox{$k=1,2,\ldots N$}\\
&& E_{k}=E_{0}+4 S J sin^{2} \frac{\pi k}{N}+2 S E_{Z}.
\label{Eq_SW}
\end{eqnarray}
Cutting the spectrum off at the OMSW states 
level is justified since above it there  
is a gap of $2 S E_{Z}$ until the lowest energy of two magnon  
spin waves. 

\noindent
The electronic and $s-d$ parts of the Hamiltonian should, in principle, 
manifest the many-body aspects of the problem (in the Kondo sense). 
In first quantization that amounts to, 
$H_{e}=-\sum_{i}\frac{d^2} {dx_{i}^2} -h_{e}
\sum_{i}s_{iz}$ and an exchange part 
$H_{s-d}= g \sum_{i} \sum_{n} \delta(x_{i}-n)
{\bf s}_{i} \cdot {\bf S}_{n}$ ($h_{e}$
is the electronic Zeeman energy). 
 The sum over $i$ runs, principally, 
on all the electrons in the wire. 
However, due to the presence of an external 
magnetic field, the ground state of the spin system is non degenerate 
and the Kondo effect is absent. Hence, 
the dynamics of the system is adequately described by the single electron
$s-d$ exchange Hamiltonian, 
\begin{eqnarray}
&&H=-\frac{d^2} {dx^2} -h_{e} 
s_{z} + g \sum_{n} \delta(x-n)
{\bf s} \cdot {\bf S}_{n}+H_{S}.
\label{Eq_Hs-d}
\end{eqnarray}
The Schr\"odinger equation is then,
\begin{eqnarray}
&& H\Psi(x;\{S_{n}\})=E\Psi(x; \{S_{n}\}),
\label{Eq_PsisS}
\end{eqnarray}
where $\Psi(x;\{S_{n}\})$ and $E$ are the total wave function 
and total energy (electron and spin system).

\noindent
{\bf The scattering problem}:
Between spins and outside the spin system, electrons
propagate as plane waves with
momenta $p_{\sigma k}= \sqrt(E-E_{k}+\sigma h_{e})$, 
(some of which might be
purely imaginary), where $\sigma=\pm \frac{1} {2}$, 
(alternatively $\sigma=\uparrow \downarrow$)
is the electron spin projection along $\hat {{\bf z}}$.
Solution of the scattering problem  
means the evaluation of transmission amplitudes 
$t_{\sigma' k' \sigma k}(E)$, 
in which $\sigma k$ are initial 
electron and spin system quantum numbers, 
while $\sigma' k'$ are the final ones. 
To extract the transmission 
amplitudes from the Schr\"odinger equation, note that   
the presence of spins at isolated points suggests 
using a transfer matrix formalism.
For $x_{n-1} < x <  x_{n}$ the total wave function is 
expanded in electron spinors $\chi_{\sigma}$ and spin states $|k>$,
\begin{eqnarray} 
\Psi(x;\{S_{n}\})=\sum_{\sigma' k'}\psi_{n \sigma' k'}(x) \chi_{\sigma'}
\otimes |k'>,
\label{Eq_Psiexpand}
\end{eqnarray}
\begin{eqnarray}
&& \psi_{n \sigma' k'}(x)=a_{n \sigma' k'}e^{i p_{\sigma' k'} (x-n)}
+b_{n \sigma' k'}e^{-ip_{\sigma' k'} (x-n)}.
\label{Eq_Psin}
\end{eqnarray}
Analogous expansion holds for $x_{n} < x < x_{n+1}$. 
The matching conditions at $x=n$ read,
\begin{eqnarray}
&& \Psi(x=n^{-};\{S_{n}\})=\Psi(x=n^{+};\{S_{n}\}),
\label{Eq_Psi1}
\end{eqnarray}
\begin{eqnarray}
&& \Psi'(x=n^{+};\{S_{n}\})-\Psi'(x=n^{-};\{S_{n}\}) \nonumber \\
&& =g {\bf s} \cdot {\bf S}_{n} 
\Psi(x=n^{-};\{S_{n}\}).
\label{Eq_Psi2}
\end{eqnarray}
Employing the expansion (\ref{Eq_Psiexpand}) and applying 
on the left the bra 
$<k|\otimes \chi_{\sigma}^{\dagger}$,
the relations (\ref{Eq_Psi1},\ref{Eq_Psi2}) imply $4(N+1)$ equations 
from which the coefficients $a_{n+1 \sigma k},
b_{n+1\sigma k}$ 
are expressible in terms of $a_{n \sigma k},b_{n \sigma k}$. 
At the end of this procedure (which is briefly detailed here), 
it yields the transfer matrix $\tau_{n}$ that carries the system 
from $n-0$ to $n+0$. To be definite, the order of coefficients 
(which determines the structure of transfer matrices) is
$({\bf a}_{n\uparrow},{\bf a}_{n\downarrow},{\bf b}_{n\uparrow},
{\bf b}_{n\downarrow}$).
The required manipulation is to evaluate 
matrix elements of four $(N+1)\times (N+1)$ matrices 
$I_{\sigma' \sigma}$ defined as operators in spin wave space,
\begin{eqnarray}
&&<k'|I_{\sigma' \sigma}|k> \equiv 
<k'|\otimes \chi^{\dagger}_{\sigma'} {\bf s} 
\cdot {\bf S}_{n} \chi_{\sigma}\otimes |k>.
\label{Eq_ISW}
\end{eqnarray}
Straightforward calculations yield, (separating $k,k'=0$ and $k,k' 
\ne 0$)
\begin{eqnarray}
&&<k'|I_{\uparrow \uparrow}|k>=\left [ \begin{array}{cc}
\frac{1}{4} & 0 \\
0 & \frac{\delta_{kk'}}{4}-\frac{e^{i(k-k')n}}{2 N} 
\end{array} \right ] , \nonumber \\
&&<k'|I_{\uparrow \downarrow}|k>=\left [ \begin{array}{cc}
0 & 0 \\
\frac{e^{-ik'n}}{2 \sqrt N} & 0 \end{array} \right ], \nonumber \\
&&I_{\downarrow \uparrow}=I_{\uparrow \downarrow}^{\dagger}, \nonumber \\
&&I_{\downarrow \downarrow}=-I_{\uparrow \uparrow}.
\label{Eq_Iexplicit}
\end{eqnarray}
The transfer matrix across $x=n$ ($\tau_{n}$) can now be 
written down, remembering that in many 
channel problems, the plane waves are normalized to have 
unit velocity. In terms of the 
$4(N+1) \times 4(N+1)$ diagonal matrix,
\begin{eqnarray}
&&p^{-\frac{1} {2}}=diag({\bf p}_{\uparrow}^{-\frac{1} {2}},
{\bf p}_{\downarrow}^{-\frac{1} {2}}
,{\bf p}_{\uparrow}^{-\frac{1} {2}}
,{\bf p}_{\downarrow}^{-\frac{1} {2}}),
\label{Eq_p12}
\end{eqnarray}
where ${\bf p}_{\sigma}
=(p_{\sigma 0},p_{\sigma 1},..p_{\sigma N})$ is the vector of 
$N+1$ channel momenta,
the result is,
\begin{eqnarray}
&& \tau_{n}={\LARGE 1}+ \nonumber \\
&& \frac{g} {2i}p^{-\frac{1} {2}}
\left [ \begin{array} {cccc}
I_{\uparrow \uparrow} & I_{\uparrow \downarrow} &
I_{\uparrow \uparrow} & I_{\uparrow \downarrow}\\
I_{\downarrow \uparrow} & I_{\downarrow \downarrow} &
I_{\downarrow \uparrow} & I_{\downarrow \downarrow} \\
-I_{\uparrow \uparrow} & -I_{\uparrow \downarrow} &
-I_{\uparrow \uparrow} & -I_{\uparrow \downarrow}\\ 
-I_{\downarrow \uparrow} & -I_{\downarrow \downarrow} &
-I_{\downarrow \uparrow} & -I_{\downarrow \downarrow} 
\end{array} \right ] p^{-\frac{1} {2}}. 
\label{Eq_tnSW}
\end{eqnarray}
Propagation between spins 
is controlled by 
the diagonal $4(N+1) \times 4(N+1)$ matrix of phases, 
\begin{eqnarray}
&&\Lambda=diag(e^{i{\bf p}_{\uparrow}},e^{i{\bf p}_{\downarrow}},
e^{-i{\bf p}_{\uparrow}},e^{-i{\bf p}_{\downarrow}}). 
\label{Eq_Lambda}
\end{eqnarray}
The transfer matrix $T_{n}$ across a unit cell (from $n-1+0$ to $n+0$)  
and the total transfer matrix $T$ are then,
$T_{n}=\Lambda \tau_{n}$ and
$T=\prod_{n=1}^{N} T_{n}$.
With some modifications required in case 
there are evanescent modes,
it is not difficult to show that in the above $4$ block 
partition of transfer matrices 
into ${\bf a}$ and ${\bf b}$
sectors, the transfer matrices 
$\tau_{n}$, $T_{n}$ and $T$ satisfy the current conservation constraint,
$\tau_{n} \sigma_{z} \tau_{n}^{\dagger}=\sigma_{z}$.
Finally, the matrix of amplitudes $t_{\sigma' \sigma}$ with 
elements $t_{\sigma' k' \sigma k}(E)$ is simply given by 
$T_{{\bf aa}}^{-1}$.  

\noindent
{\bf Conductance and spin transfer coefficients}:
An experimentally relevant question is the following: 
Applying a small voltage $V$ across the wire between 
$0$ and $N+1$ and letting a
unit flux of electrons at Fermi energy $\varepsilon_{F}$
and spin component $\sigma$ reach
it from the left,  what is the current $I_{\sigma' \sigma}$ 
of electrons with 
spin $\sigma'$ in the system? The ratio 
$G_{\sigma' \sigma}(\varepsilon_{F}) \equiv I_{\sigma' \sigma}/V$
is the corresponding STC. 
The conductance per spin is 
$G_{\sigma}=\sum_{\sigma'}G_{\sigma' \sigma}$ and the conductance 
is $G=\sum_{\sigma} G_{\sigma}$. 
At zero temperature (and for infinitesimal voltage drop $V$) 
the system is in its ground state and
only elastic scattering (of the electrons) is allowed. 
Due to spin conservation, processes involving spin flip 
are inelastic and hence are forbidden. Thus, the 
quantities to be calculated at zero temperature and infinitesimal 
voltage difference are,
\begin{eqnarray}
&&G_{\sigma \sigma}(\varepsilon_{F})=
|t_{\sigma 0 \sigma 0}(\varepsilon_{F}+E_{0})|^{2},
\label{Eq_Gs's}
\end{eqnarray}
where $\varepsilon_{F}+E_{0}=E$ is the total energy of the system. 
Evidently, $G_{\uparrow \uparrow}$ is trivial since it involves 
only the ground state $|0>$ of the spin system. Coupling 
terms to OMSW, 
$<k \ne 0|\otimes \chi_{\sigma '}^{\dagger} {\bf s} \cdot {\bf S}_{n}
\chi_{\uparrow} \otimes |0>=0$ since the initial state 
$\chi_{\uparrow}\otimes |0>$ is stretched with total 
spin $N/2+1/2$ 
while the final state is not. 
Therefore, the quantum spins are 
replaced by their averages $<0|{\bf S}_{n}|0>=S_{nz}$ and 
spin up incoming electrons are  
scattered by a sequence of 
$N$ identical delta function potentials of strength $g/4$. 
On the other hand,  $G_{\downarrow \downarrow}$ involves 
virtual OMSW excitations and hence it is non trivial. 
For examples, it gets second order contributions 
 $<0|\otimes \chi_{\downarrow}^{\dagger} {\bf s} \cdot {\bf S}_{m}
\chi_{\uparrow}\otimes |k>
<k|\otimes \chi_{\uparrow}^{\dagger} {\bf s} \cdot {\bf S}_{n}
\chi_{\downarrow}\otimes |0>$. \\
Of special interest are spin flip processes.
If the potential drop $eV$ exceeds 
 the gap between the ground state energy $E_{0}$ and 
the lowest OMSW energy (that is, $eV \ge 2 S E_{Z}$), then the 
process $\uparrow \leftarrow \downarrow $ is energetically 
feasible even at $T=0$. The electron stays at the Fermi energy and the 
spin system is heated from $E_{0}$ to $E_{1}$, the energy being 
supplied from the external voltage.  Assuming that linear response is still 
valid, the corresponding STC is
$G_{\uparrow  \downarrow }(\varepsilon_{F})=
|t_{\uparrow 1 \downarrow 0}(\varepsilon_{F}+E_{0})|^{2}$.
After the system is excited from the ground state to the 
lowest OMSW state, the process $\uparrow \leftarrow \downarrow $
is blocked. It can occur again only if the spin system returns to its 
ground state. This can happen for example by the process 
$\downarrow \leftarrow \uparrow $ (which, by spin conservation, 
is allowed if the spin system is in a OMSW state but is otherwise 
forbidden on the ground state).
Finally, consider the spin-flip process encoded in
$G_{\downarrow \uparrow}(\varepsilon_{F})=
|t_{\downarrow 0 \uparrow k}(\varepsilon_{F}+E_{k})|^{2}$.
This can happen only at $T>0$ where the spin system
contains also a OMSW component. At the end, the spin system cools down to
the ground state, but regains the excited component energy from the
heat bath. \\
{\bf Possible experimental realization}: 
Out of the four parameters $g$, $J$, ${\bf H}$ and 
$\varepsilon_{F}$, the last two are the ones which are easiest 
to handle and control. However, 
the strength of the magnetic field is somewhat constrained 
by the condition  $E_{Z} > JS$ justifying 
projection on the OMSW subspace. The $s-d$ coupling
$g$ is material dependent and can have either sign but in general 
$|g| \gg |J|$. The atomic and electronic  
Zeeman energies are similar in magnitude, $E_{Z} \sim h_{e}$.  
Thus, one should concieve a material consisting 
of itinerant electrons and localized magnetic moments 
in which the Fermi energy can be tuned by a sutable gate. 
Possible candidates are
magnetic semiconductors ({\em e.g.} $FeSi$) or strongly 
correlated semimetals such as $CeP$ and $CeSb$. \\
{\bf Results}:
Calculations are performed on a chain of $N=12$ spins, 
using $JS=0.06$, and taking $g=\pm 1$ for 
the electron-atom $s-d$ exchange strength. The atomic and 
electronic Zeeman energies are taken to be $E_{Z}=h_{e}=0.2$. The 
Fermi energy is taken with reference to the ground state of the 
spin system in units of $|J|S$.
The results at $T=0$ are presented in figures \ref{fig1}, \ref{fig2}, 
and \ref{fig3} below.
\begin{figure}[htb]
\includegraphics[width=0.35\textwidth,keepaspectratio]{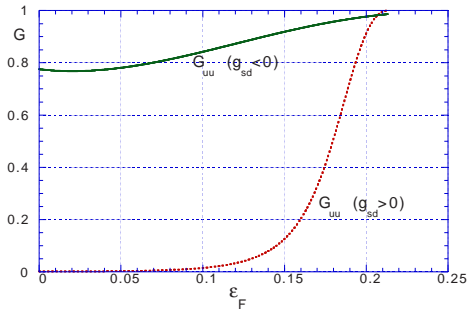}
\caption
{Non-flip spin transfer coefficient
$G_{\uparrow \uparrow }(\varepsilon_{F})$ (equation \ref{Eq_Gs's}) 
for electron transmission through a
chain consisting of $N=12$ spins 
prepared at the ground state $|0>$. Solid (dashed) lines correspond to 
$g=-1$, a ferromagnetic ($g=1$, an antiferromagnetic) 
$s-d$ exchange coupling. 
Other parameters are $J=0.12$, $E_{Z}=h_{e}=0.2$. }
\label{fig1}
\end{figure}
\begin{figure}[htb]
\includegraphics[width=0.35\textwidth,keepaspectratio]
{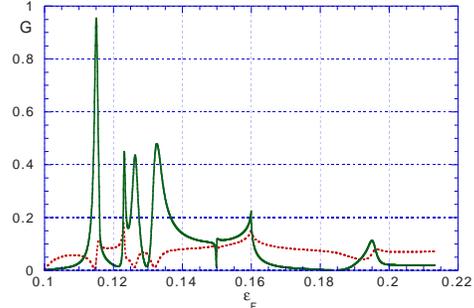}
\caption
{Non-flip spin transfer coefficient
$G_{\downarrow \downarrow }(\varepsilon_{F})$ (solid line) 
and spin flip transfer coefficient 
$G_{\uparrow \downarrow }(\varepsilon_{F})$ (dashed line) 
for a ferromagnetic $s-d$ exchange coupling $g=-1$. 
Other specifications are as in figure \ref{fig1}.}
\label{fig2}
\end{figure}
\begin{figure}[htb]
\includegraphics[width=0.35\textwidth,keepaspectratio]
{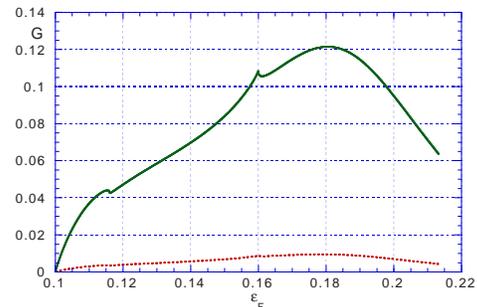}
\caption
{Same as figure \ref{fig2} for $g=1$ 
(an anti ferromagnetic $s-d$ exchange coupling).}
\label{fig3}
\end{figure}
As asserted above, the behavior of $G_{\uparrow \uparrow}$ (figure
\ref{fig1}) is somewhat expected since only 
the $s_{z}S_{nz}$ interaction is effective. For a ferromagnetic $s-d$
coupling 
the electron encounters 
a periodic potential composed of attractive $\delta$ functions 
while for an antiferromagnetic exchange the potentials 
are repulsive.  
This leads
to a band structure determined by,
$|cosk_{0} + \frac {g} {k_{0}} sink_{0}| \le 1$,
where $k_{0}=\sqrt\varepsilon_{F}$. 
It is easy to see that the above range of 
$\varepsilon_{F}$ 
is inside the band. \\
The patterns of the STC 
$G_{\downarrow \downarrow}$ and
$G_{\uparrow \downarrow}$ (figures \ref{fig2} for 
$g=-1$ and \ref{fig3} for $g=1$) 
have a much richer content. In particular, 
for a ferromagnetic exchange coupling (figure \ref{fig2}) they display 
a series of very narrow resonances. It is quite remarkable that 
in a magnetic field and with ferromagnetic $s-d$ coupling, still, 
$G_{\downarrow \downarrow}$ approaches unity at resonance.
These resonances are novel, that is, they 
are not the usual ones encountered in
resonance tunneling, since there is neither potential 
scattering here nor double barrier. 
As discussed above, the pertinent processes 
 involve virtual 
excitations of the OMSW, and hence, it reflects the 
structure of the electron spin chain system. Unlike potential 
scattering in which the scattering is effected upon 
static (quenched) impurities, we encounter here scattering
from a quantum system with an internal structure and many 
non-degenerate levels (the case of degenerate levels 
is exemplified by the Kondo effect). 
It is also noticed that these resonances appear only for 
a ferromagnetic $s-d$ coupling $g<0$. 
Some light can be shed on
 these points by inspecting the corresponding bound state 
problem for a {\em closed} system. Namely, 
the one dimensional wire is closed into a ring 
of length $N$ and the bound state energies 
(pertaining to the {\em total} system, electron and spin chain) 
can be computed by solving the equation $det[T(E)-1]=0$ 
(in order to avoid orbital effects, the magnetic field is 
tuned  to yield an nteger number of flux quanta through the 
ring). 
While there is no simple relation between an open wire 
conductance rsonances and bound states on a closed wire,
it is nevertheless instructive to  
notice that the bound state energies in the
ferromagnetic case are lower than those in the 
antiferromagnetic case. In fact, the latters fall mostly outside 
the range of Fermi energies examined in these simulations. \\
At finite temperature, the spin system can be 
activated to the lowest OMSW (which, unlike higher OMSW, 
is not degenerate). The process 
$\downarrow 1 \leftarrow \uparrow 0$ is then 
feasible and the  relevant STC
$G_{\downarrow \uparrow}$ is depicted in figure \ref{fig4}.  
\begin{figure}[htb]
\includegraphics[width=0.35\textwidth,keepaspectratio]
{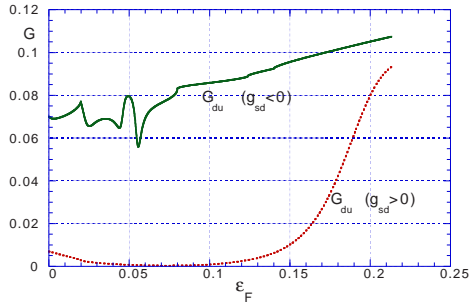}
\caption
{The spin transfer coefficient 
$G_{\downarrow  \uparrow }(\varepsilon_{F})$  
for a spin system
prepared at the first OMSW state. The value of the $s-d$ 
exchange constant is $g=\mp 1$ (solid and dashed lines).
Other parameters are 
as in figure \ref{fig1}.}
\label{fig4}
\end{figure}
\noindent 
It mainly reflects 
the coupling between the 
ground and lowest OMSW 
state of the spin system. Higher OMSW are virtually excited in higher 
orders through non-flip processes, {\em e.g.} 
$(\downarrow 0) \leftarrow (\uparrow k \ge 1) \leftarrow (\uparrow 1).$
The resonant pattern prevails also 
in this case. Note that the actual value of the STC should be 
reduced by the activation factor $e^{-\beta (E_{1}-E_{0})}$. \\
{\bf Conclusions}: The formalism for studying electron transport 
in a quantum spin chain is developed. Application of an  
external magnetic field removes the Kondo problem as it
lifts the degeneracy of spin states. (For infinite chains 
it also restores long range order which is otherwise 
absent due to the Mermin - Wagner theorem). Spin transfer 
coefficients $G_{\sigma' \sigma}$ are evaluated and 
analyzed as a function of the Fermi energy. 
Non-flip processes $\uparrow \leftarrow \uparrow$ 
and $\downarrow \leftarrow \downarrow $ are elastic.
The structure of $G_{\uparrow \uparrow }$ 
is trivial since the problem is 
equivalent to the scattering of polarized electrons on a 
sequence of delta function potentials. On the other hand, 
the behavior of $G_{\downarrow \downarrow}$ is 
much richer.  For a ferromagnetic $s-d$ exchange coupling 
it displays a novel resonant pattern 
which is distinct from the one encountered in 
potential or double barrier transmission. That is, 
it reflects the complex structure of the electron spin-chain system. 
Spin-flip processes  are inelastic. The transition 
$\uparrow \leftarrow \downarrow $ is endothermic, and is 
 feasible at $T=0$ if the potential difference across 
the spin system overcomes the energy gap between the 
ground and 
lowest OMSW states. The transition 
$\downarrow \leftarrow \uparrow $  is exothermic 
and feasible at $T>0$ if 
OMSW are activated. Although spin flip coefficients 
are small relative to  
non spin flip ones, they are still sizable. 
 In passing, we 
note that there are interesting theoretical works discussing the 
pertinent many body problem \cite{Fujimoto} \cite{Zachar}.\\ 
{\bf Acknowledgements}: We thank M. Yamanaka, T. Koma, 
E. Kogan and K. Kikoin for 
helpful comments. Y. A. acknowledges
supports by DIP German Israel Cooperation, 
Israeli SF grants {\em Center of Excellence}, 
{\em Many Body Effects in Resonance Tunneling} and 
US-Israel BSF {\em Dynamical Instabilities in quantum dots}. 
Y.T. acknowledges support from the
NEDO joint research program (NTDP-98).

\end{multicols}
\end{document}